\pdfoutput=1
\documentclass[12pt,a4paper]{article}

\usepackage{ifthen} 
\newboolean{pdflatex}
\setboolean{pdflatex}{true} 

\newboolean{articletitles}
\setboolean{articletitles}{true} 

\newboolean{uprightparticles}
\setboolean{uprightparticles}{false} 

\def\paperauthors{LHCb collaboration} 
\def\paperasciititle{Evidence of a JpsiKS structure in B0->Jpsi phi KS decays} 
\def\papertitle{Evidence of a $\jpsi\KS$ structure in $\Bz\to\jpsi\phi\KS$ decays} 
\def\paperkeywords{{High Energy Physics}, {LHCb}} 
\def\papercopyright{\the\year\ CERN for the benefit of the LHCb collaboration} 
\def\paperlicence{CC BY 4.0 licence}
\def\paperlicenceurl{https://creativecommons.org/licenses/by/4.0/}

\usepackage[top=1in, bottom=1.25in, left=1in, right=1in]{geometry}

\usepackage{microtype}
\usepackage{lineno}  
\usepackage{xspace} 
\usepackage{caption} 

\usepackage{graphicx}  
\usepackage{color}
\usepackage{colortbl}

\usepackage{amsmath} 
\usepackage{amssymb}
\usepackage{amsfonts}
\usepackage{upgreek} 

\newcommand*\patchAmsMathEnvironmentForLineno[1]{%
\expandafter\let\csname old#1\expandafter\endcsname\csname #1\endcsname
\expandafter\let\csname oldend#1\expandafter\endcsname\csname
end#1\endcsname
 \renewenvironment{#1}%
   {\linenomath\csname old#1\endcsname}%
   {\csname oldend#1\endcsname\endlinenomath}%
}
\newcommand*\patchBothAmsMathEnvironmentsForLineno[1]{%
  \patchAmsMathEnvironmentForLineno{#1}%
  \patchAmsMathEnvironmentForLineno{#1*}%
}
\AtBeginDocument{%
\patchBothAmsMathEnvironmentsForLineno{equation}%
\patchBothAmsMathEnvironmentsForLineno{align}%
\patchBothAmsMathEnvironmentsForLineno{flalign}%
\patchBothAmsMathEnvironmentsForLineno{alignat}%
\patchBothAmsMathEnvironmentsForLineno{gather}%
\patchBothAmsMathEnvironmentsForLineno{multline}%
\patchBothAmsMathEnvironmentsForLineno{eqnarray}%
}

\usepackage{hyperxmp}

\usepackage[pdftex,
            pdfauthor={\paperauthors},
            pdftitle={\paperasciititle},
            pdfkeywords={\paperkeywords},
            pdfcopyright={Copyright (C) \papercopyright},
            pdflicenseurl={\paperlicenceurl}]{hyperref}

\usepackage[colorinlistoftodos,textsize=scriptsize]{todonotes}

\usepackage[bottom,flushmargin,hang,multiple]{footmisc}

\usepackage[all]{hypcap} 

\usepackage{xspace} 
\usepackage{upgreek}

\def\lhcb   {\mbox{LHCb}\xspace}

\def\besiii {\mbox{BESIII}\xspace}

\def\MagUp {\mbox{\em Mag\kern -0.05em Up}\xspace}

\ifthenelse{\boolean{uprightparticles}}%
{

 \def\Pmu         {\ensuremath{\upmu}\xspace}

 \def\Ppi         {\ensuremath{\uppi}\xspace}

 \def\Ppsi        {\ensuremath{\uppsi}\xspace}

 \def\PDelta      {\ensuremath{\Delta}\xspace}                 
 \def\PXi         {\ensuremath{\Xi}\xspace}                 
 \def\PLambda     {\ensuremath{\Lambda}\xspace}                 
 \def\PSigma      {\ensuremath{\Sigma}\xspace}                 
 \def\POmega      {\ensuremath{\Omega}\xspace}                 
 \def\PUpsilon    {\ensuremath{\Upsilon}\xspace}
 \let\oldPi\Pi
 \def\PPi         {\ensuremath{\oldPi}\xspace}

 \def\PB      {\ensuremath{\mathrm{B}}\xspace}                 
                  
 \def\PD      {\ensuremath{\mathrm{D}}\xspace}

 \def\PJ      {\ensuremath{\mathrm{J}}\xspace}                 
 \def\PK      {\ensuremath{\mathrm{K}}\xspace}

 \def\Pb      {\ensuremath{\mathrm{b}}\xspace}                 
 \def\Pc      {\ensuremath{\mathrm{c}}\xspace}                 
 \def\Pd      {\ensuremath{\mathrm{d}}\xspace}

 \def\Pi      {\ensuremath{\mathrm{i}}\xspace}

 \def\Pp      {\ensuremath{\mathrm{p}}\xspace}

 \def\Ps      {\ensuremath{\mathrm{s}}\xspace}                 
                  
 \def\Pu      {\ensuremath{\mathrm{u}}\xspace}

 \def\thebaroffset{0.0em}
}
{

 \def\Pmu         {\ensuremath{\mu}\xspace}

 \def\Ppi         {\ensuremath{\pi}\xspace}

 \def\Ppsi        {\ensuremath{\psi}\xspace}                 
                  
 \mathchardef\PDelta="7101
 \mathchardef\PXi="7104
 \mathchardef\PLambda="7103
 \mathchardef\PSigma="7106
 \mathchardef\POmega="710A
 \mathchardef\PUpsilon="7107
 \mathchardef\PPi="7105
                  
 \def\PB      {\ensuremath{B}\xspace}                 
                  
 \def\PD      {\ensuremath{D}\xspace}

 \def\PJ      {\ensuremath{J}\xspace}                 
 \def\PK      {\ensuremath{K}\xspace}

 \def\Pb      {\ensuremath{b}\xspace}                 
 \def\Pc      {\ensuremath{c}\xspace}                 
 \def\Pd      {\ensuremath{d}\xspace}

 \def\Pi      {\ensuremath{i}\xspace}

 \def\Pp      {\ensuremath{p}\xspace}

 \def\Ps      {\ensuremath{s}\xspace}                 
                  
 \def\Pu      {\ensuremath{u}\xspace}

 \def\thebaroffset{0.18em}
}
\newcommand{\offsetoverline}[2][\thebaroffset]{\kern #1\overline{\kern -#1 #2}}%

\makeatletter
\ifcase \@ptsize \relax
  \newcommand{\miniscule}{\@setfontsize\miniscule{4}{5}}
\or
  \newcommand{\miniscule}{\@setfontsize\miniscule{5}{6}}
\or
  \newcommand{\miniscule}{\@setfontsize\miniscule{5}{6}}
\fi
\makeatother

\DeclareRobustCommand{\optbar}[1]{\shortstack{{\miniscule (\rule[.5ex]{1.25em}{.18mm})}
  \\ [-.7ex] $#1$}}

\def\mup        {{\ensuremath{\Pmu^+}}\xspace}
\def\mun        {{\ensuremath{\Pmu^-}}\xspace} 

\def\uquark    {{\ensuremath{\Pu}}\xspace}

\def\dquark    {{\ensuremath{\Pd}}\xspace}

\def\squark    {{\ensuremath{\Ps}}\xspace}
\def\squarkbar {{\ensuremath{\overline \squark}}\xspace}

\def\cquark    {{\ensuremath{\Pc}}\xspace}
\def\cquarkbar {{\ensuremath{\overline \cquark}}\xspace}

\def\bquark    {{\ensuremath{\Pb}}\xspace}

\def\pion   {{\ensuremath{\Ppi}}\xspace}

\def\pimp   {{\ensuremath{\pion^\mp}}\xspace}

\def\kaon    {{\ensuremath{\PK}}\xspace}
\def\Kbar    {{\ensuremath{\offsetoverline{\PK}}}\xspace}
\def\Kb      {{\ensuremath{\Kbar}}\xspace}
\def\KorKbar {\kern \thebaroffset\optbar{\kern -\thebaroffset \PK}{}\xspace}

\def\Kp      {{\ensuremath{\kaon^+}}\xspace}
\def\Km      {{\ensuremath{\kaon^-}}\xspace}
\def\Kpm     {{\ensuremath{\kaon^\pm}}\xspace}

\def\KS      {{\ensuremath{\kaon^0_{\mathrm{S}}}}\xspace}

\def\Kstar   {{\ensuremath{\kaon^*}}\xspace}

\def\Dbar    {{\ensuremath{\offsetoverline{\PD}}}\xspace}
\def\D       {{\ensuremath{\PD}}\xspace}

\def\DorDbar {\kern \thebaroffset\optbar{\kern -\thebaroffset \PD}\xspace}

\def\Dzb     {{\ensuremath{\Dbar{}^0}}\xspace}
\def\Dp      {{\ensuremath{\D^+}}\xspace}
\def\Dm      {{\ensuremath{\D^-}}\xspace}

\def\DpDm    {\ensuremath{\Dp {\kern -0.16em \Dm}}\xspace}

\def\Dstarb  {{\ensuremath{\Dbar{}^*}}\xspace}

\def\Dstarzb {{\ensuremath{\Dbar{}^{*0}}}\xspace}

\def\Dstarm  {{\ensuremath{\D^{*-}}}\xspace}

\def\Dsp     {{\ensuremath{\D^+_\squark}}\xspace}
\def\Dsm     {{\ensuremath{\D^-_\squark}}\xspace}

\def\Dssp    {{\ensuremath{\D^{*+}_\squark}}\xspace}

\def\B       {{\ensuremath{\PB}}\xspace}
\def\Bbar    {{\ensuremath{\offsetoverline{\PB}}}\xspace}

\def\BorBbar {\kern \thebaroffset\optbar{\kern -\thebaroffset \PB}\xspace}
\def\Bz      {{\ensuremath{\B^0}}\xspace}
\def\Bzb     {{\ensuremath{\Bbar{}^0}}\xspace}
\def\Bd      {{\ensuremath{\B^0}}\xspace}

\def\BdorBdbar {\kern \thebaroffset\optbar{\kern -\thebaroffset \Bd}\xspace}
\def\Bu      {{\ensuremath{\B^+}}\xspace}
\def\Bub     {{\ensuremath{\B^-}}\xspace}
\def\Bp      {{\ensuremath{\Bu}}\xspace}
\def\Bm      {{\ensuremath{\Bub}}\xspace}

\def\Bs      {{\ensuremath{\B^0_\squark}}\xspace}

\def\BsorBsbar {\kern \thebaroffset\optbar{\kern -\thebaroffset \Bs}\xspace}

\def\jpsi     {{\ensuremath{{\PJ\mskip -3mu/\mskip -2mu\Ppsi}}}\xspace}

\def\Y#1S{\ensuremath{\PUpsilon{(#1S)}}\xspace}

\def\proton      {{\ensuremath{\Pp}}\xspace}
\def\antiproton  {{\ensuremath{\overline \proton}}\xspace}

\def\LorLbar     {\kern \thebaroffset\optbar{\kern -\thebaroffset \PLambda}\xspace}
\def\Lambdares   {{\ensuremath{\PLambda}}\xspace}

\newcommand{\decay}[2]{\ensuremath{#1\!\to #2}\xspace} 

\def\to                 {\ensuremath{\rightarrow}\xspace}

\def\AT#1     {\ensuremath{A_{\mathrm{T}}^{#1}}\xspace}           

\def\C#1      {\ensuremath{\mathcal{C}_{#1}}\xspace}                       
\def\Cp#1     {\ensuremath{\mathcal{C}_{#1}^{'}}\xspace}                    
\def\Ceff#1   {\ensuremath{\mathcal{C}_{#1}^{\mathrm{(eff)}}}\xspace}        
\def\Cpeff#1  {\ensuremath{\mathcal{C}_{#1}^{'\mathrm{(eff)}}}\xspace}       
\def\Ope#1    {\ensuremath{\mathcal{O}_{#1}}\xspace}                       
\def\Opep#1   {\ensuremath{\mathcal{O}_{#1}^{'}}\xspace}                    

\newcommand{\aunit}[1]{\ensuremath{\text{\,#1}}}       

\newcommand{\tev}{\aunit{Te\kern -0.1em V}\xspace}
\newcommand{\gev}{\aunit{Ge\kern -0.1em V}\xspace}
\newcommand{\mev}{\aunit{Me\kern -0.1em V}\xspace}
\newcommand{\kev}{\aunit{ke\kern -0.1em V}\xspace}
\newcommand{\ev}{\aunit{e\kern -0.1em V}\xspace}
 
\newcommand{\mevc}{\ensuremath{\aunit{Me\kern -0.1em V\!/}c}\xspace}
\newcommand{\gevc}{\ensuremath{\aunit{Ge\kern -0.1em V\!/}c}\xspace}
\newcommand{\mevcc}{\ensuremath{\aunit{Me\kern -0.1em V\!/}c^2}\xspace}
\newcommand{\gevcc}{\ensuremath{\aunit{Ge\kern -0.1em V\!/}c^2}\xspace}

\def\fb   {\ensuremath{\aunit{fb}}\xspace}
\def\invfb   {\ensuremath{\fb^{-1}}\xspace}

\newcommand{\chisq}{\ensuremath{\chi^2}\xspace}

\def\gsim{{~\raise.15em\hbox{$>$}\kern-.85em
          \lower.35em\hbox{$\sim$}~}\xspace}
\def\lsim{{~\raise.15em\hbox{$<$}\kern-.85em
          \lower.35em\hbox{$\sim$}~}\xspace}

\def\evtgen     {\mbox{\textsc{EvtGen}}\xspace}

\def\geant      {\mbox{\textsc{Geant4}}\xspace}

\def\pythia     {\mbox{\textsc{Pythia}}\xspace}

\def\tell1  {TELL1\xspace}
\def\ukl1   {UKL1\xspace}

\newcommand{\phz}{\phantom{0}}

\newcommand{\lhcborcid}[1]{\href{https://orcid.org/#1}{\hspace*{0.1em}\raisebox{-0.45ex}{\includegraphics[width=1em]{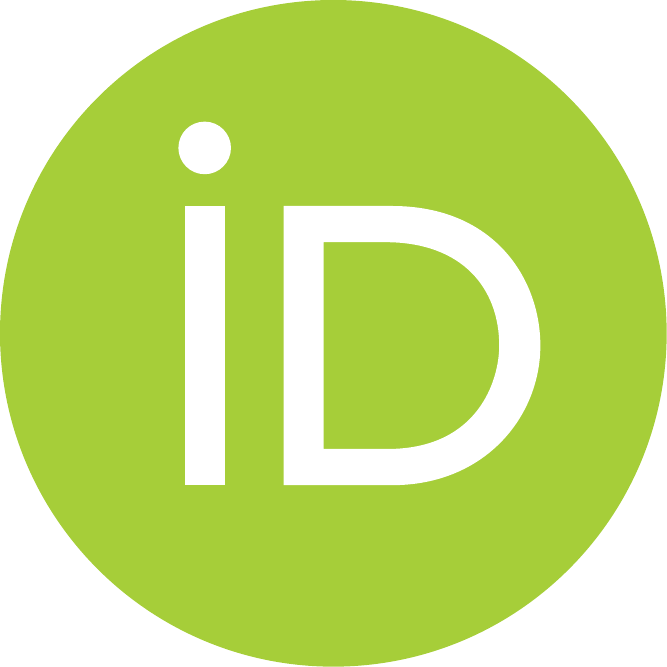}}}}

\usepackage{cite} 
\usepackage{mciteplus}

\def\Bpdecay{\mbox{\ensuremath{\Bp\to\jpsi\phi\Kp}}\xspace}
\def\Bzdecay{\mbox{\ensuremath{\Bz\to\jpsi\phi\KS}}\xspace}
\def\zcsAll{\ensuremath{T_{\psi s1}}\xspace}
\def\zcszAll{\ensuremath{T_{\psi s}^{ 0}}\xspace}
\def\zcspAll{\ensuremath{T_{\psi s}^{ +}}\xspace}
\def\zcsz{\ensuremath{T_{\psi s1}^{\theta}(4000)^0}\xspace}
\def\zcsOne{\ensuremath{T_{\psi s1}^{\theta}(4000)}\xspace}
\def\zcsTwo{\ensuremath{T_{\psi s1}(4220)}\xspace}
\def\zcsp{\ensuremath{T_{\psi s1}^{\theta}(4000)^+}\xspace}
\def\zcspa{\ensuremath{T_{\psi s1}(4220)^+}\xspace}
\def\zcspbes{\ensuremath{Z_{cs}(3985)^+}\xspace}
\def\zcszbes{\ensuremath{Z_{cs}(3985)^0}\xspace}

\usepackage{longtable} 
\usepackage{multirow}
\usepackage{subfigure}

\begin{document}

\renewcommand{\thefootnote}{\fnsymbol{footnote}}
\setcounter{footnote}{1}

\begin{titlepage}
\pagenumbering{roman}

\vspace*{-1.5cm}
\centerline{\large EUROPEAN ORGANIZATION FOR NUCLEAR RESEARCH (CERN)}
\vspace*{1.5cm}
\noindent
\begin{tabular*}{\linewidth}{lc@{\extracolsep{\fill}}r@{\extracolsep{0pt}}}
\ifthenelse{\boolean{pdflatex}}
{\vspace*{-1.5cm}\mbox{\!\!\!\includegraphics[width=.14\textwidth]{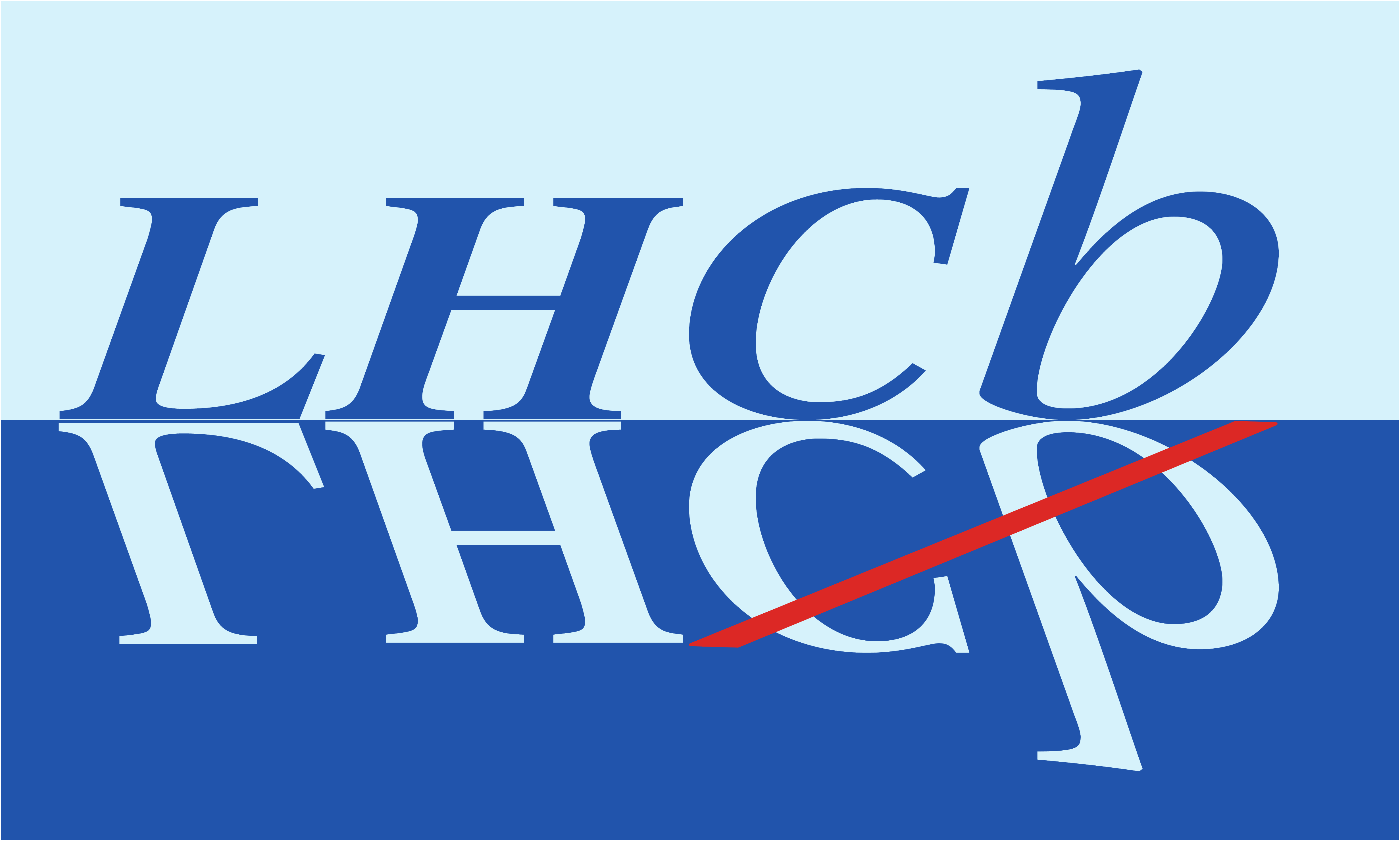}} & &}%
{\vspace*{-1.2cm}\mbox{\!\!\!\includegraphics[width=.12\textwidth]{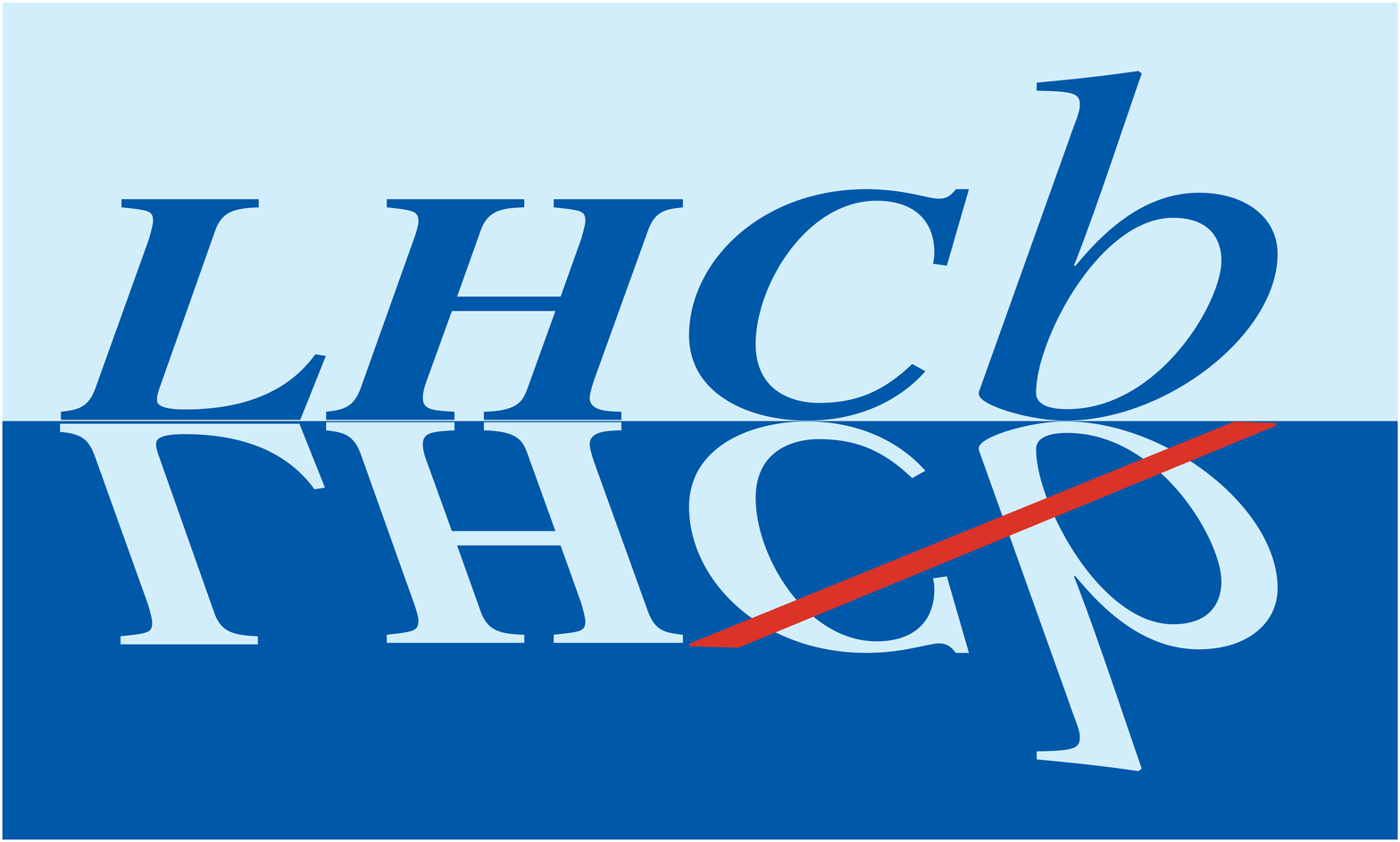}} & &}%
\\
    & & CERN-EP-2022-258 \\  
    & & LHCb-PAPER-2022-040 \\  
    & & 7 October 2023 \\ 
    & & \\
\end{tabular*}

\vspace*{4.0cm}

{\normalfont\bfseries\boldmath\huge
\begin{center}
    \papertitle 
\end{center}
}

\vspace*{1.0cm}

\begin{center}
\paperauthors\footnote{Authors are listed at the end of this Letter.}
\end{center}

\vspace{\fill}

\begin{abstract}
    \noindent
    An amplitude analysis of $\Bz \to \jpsi \phi \KS$ decays is performed using proton-proton collision data,  corresponding to an integrated luminosity of 9\invfb, collected with the \lhcb detector at centre-of-mass energies of 7, 8 and 13\tev.
    Evidence with a significance of $4.0$ standard deviations of a structure in the $\jpsi\KS$ system, named \zcsz, is seen, with its mass and width measured to be $3991 \,_{-10}^{+12} \,^{+\phz9}_{-17} \mevcc$ and $105 \,_{-25}^{+29}  \,^{+17}_{-23} \mev$, respectively, where the first uncertainty is statistical and the second systematic.
    The \zcsz state is likely to be the isospin partner of the \zcsp state, previously observed in the $\jpsi \Kp$ system of the $\Bpdecay$ decay.
    When isospin symmetry for the charged and neutral $\zcsOne$ states is assumed, the signal significance increases to $5.4$ standard deviations.  

\end{abstract}

\vspace*{2.0cm}

\begin{center}
    Published in Phys. Rev. Lett. 131 (2023) 131901 
\end{center}

\vspace{\fill}

{\footnotesize 
\centerline{\copyright~\papercopyright. \href{\paperlicenceurl}{\paperlicence}.}}
\vspace*{2mm}

\end{titlepage}

\newpage
\setcounter{page}{2}
\mbox{~}

\renewcommand{\thefootnote}{\arabic{footnote}}
\setcounter{footnote}{0}

\cleardoublepage


\pagestyle{plain} 
\setcounter{page}{1}
\pagenumbering{arabic}


Hadrons formed by more than three quarks, namely exotic hadrons, can have complex colour and flavour structures, and studies of the structures shed light on the nature of the strong interaction.
Since the discovery of the $\chi_{c1}(3872)$~\cite{Belle:2003nnu}, several states compatible with a four or five quark composition have been observed~\cite{PDG2022}, including the fully charmed tetraquark states~\cite{LHCb-PAPER-2020-011,ATLAS:2022hhx,CMS:2022yhl} and the pentaquark states~\cite{LHCb-PAPER-2019-014,LHCB-PAPER-2022-031}.
In 2020, the \besiii collaboration reported the observation of the \zcspbes state  in the $\Dsp \Dstarzb$ and $\Dssp\Dzb$ mass spectra~\cite{BESIII:2020qkh}.\footnote{Charge conjugation is implied throughout the Letter unless specified otherwise.}
Two similar states, \zcsp and \zcspa, were observed in \mbox{$\decay{\Bp}{\jpsi\phi\Kp}$} decays by the \lhcb experiment~\cite{LHCb-PAPER-2020-044}.\footnote{The new exotic hadron naming convention proposed by the \lhcb collaboration~\cite{Gershon:2022xnn} is applied.  The \zcsp, and \zcspa states are named $Z_{cs}(4000)^+$, and $Z_{cs}(4220)^+$ in the original papers~\cite{BESIII:2020qkh,LHCb-PAPER-2020-044}, respectively. The superscript $\theta$ refers to 1/2 isospin and positive parity.} 
The observations of the \zcsp, \zcspa, and \zcspbes states have stimulated many theoretical studies to interpret their internal structures, such as hadronic molecules~\cite{Meng:2020ihj}, compact tetraquarks~\cite{Maiani:2021tri}, or threshold effects~\cite{Ikeno:2021mcb}.

Searching for $\zcszAll$, the isospin partner of the $\zcspAll$, is an important step in identifying the full SU(3) nonet that involves these states.  
The quark contents for the $\zcspAll$ and $\zcszAll$ states are $\cquark\cquarkbar\uquark\squarkbar$ and $\cquark\cquarkbar\dquark\squarkbar$, respectively. 
Recently, the \besiii experiment reported evidence for the \zcszbes state ($\cquark\cquarkbar\dquark\squarkbar$) in the $\Dsp \Dstarm$ and $\Dssp\Dm$ mass spectra~\cite{BESIII:2022qzr}.
The $\Bzdecay$ decay~\cite{CLEO:1999alr,BaBar:2003mjy,BaBar:2014wwp} is the ideal candidate to search for such \zcszAll states.
The quark level processes of the $\Bzdecay$ decay are similar to those of the $\Bpdecay$ decay, as shown in Fig.~1(c,d) of Ref.~\cite{BaBar:2003mjy}, and the two decays are related by isospin symmetry.

In this Letter, evidence of a $\jpsi\KS$ structure is reported by making use of  an amplitude analysis of \mbox{$\Bd \to \jpsi \phi \KS$} decays. 
The analysis is based on proton-proton ($pp$) collision data, collected with the \lhcb detector at centre-of-mass energies of $7$, $8$, and $13\tev$, corresponding to integrated luminosities of 1, 2, and 6\invfb, respectively.

The LHCb detector is a single-arm forward spectrometer covering the pseudorapidity range \mbox{$2 < \eta < 5$}, 
described in detail in Refs.~\cite{LHCb-DP-2008-001,LHCb-DP-2014-002}. 
Simulated samples of the signal decays are used to estimate the effect of reconstruction and event selections. 
The samples are generated with \pythia~\cite{Sjostrand:2007gs,*Sjostrand:2006za}, \evtgen~\cite{Lange:2001uf}, and the \geant toolkit~\cite{Allison:2006ve, *Agostinelli:2002hh} as described in Ref.~\cite{LHCb-PROC-2011-006}.

For the $\Bzdecay$ decays, the \jpsi, $\phi$, and \KS candidates are reconstructed by combining two oppositely charged muons, kaons, and pions, respectively, forming vertices detached from any primary vertex (PV).
The \Bd candidate is reconstructed by combining the \jpsi, $\phi$, and \KS candidates, and the resulting decay vertex must be of high quality. 
In order to improve the resolution of the mass of the $\Bd$ candidates (referred to as $m_{\jpsi\phi\KS}$), a kinematic constraint~\cite{Hulsbergen_2005} is applied, where the momenta of the final-state particles are recalculated constraining the masses of the $\jpsi$ and $\KS$ candidates to their nominal values, and the momentum of $\Bz$ candidate pointing back to its associated PV. 
The associated PV is the one with the smallest impact parameter $\chisq$, defined as the difference in the vertex-fit $\chisq$ of a given PV with and without the particle under consideration.

The $\Bz$ candidates are selected by a multivariate classifier based on a multilayer perceptron~\cite{mcculloch1990logical,hopfield1982neural}. The selection is similar to that used in Ref.~\cite{LHCb-PAPER-2020-044}, with the additional requirement on the significance of the flight distance of the $\KS$ from the $\Bz$ decay vertex. 
Only candidates with the mass of the $\Kp\Km$ system within $\pm15\mev$ around the nominal $\phi$ mass~\cite{PDG2022} are kept, where the mass resolution is about $0.7 \mev$.\footnote{Natural units with $\hbar = c=1$ are used throughout the Letter.}
The multivariate classifier uses eleven variables related to the decay-chain topology, particle transverse momentum, 
vertex fit quality, and charged particle identification information. 
The selection criterion on the classifier response is chosen by maximising the figure of merit, \mbox{$\mathcal{S}^2/(\mathcal{S}+\mathcal{B})^{3/2}$}~\cite{LHCb-PAPER-2018-045}, where $\mathcal{S}$ and $\mathcal{B}$ are the yields of signal and background in the signal region, respectively. 
The signal region in $m_{\jpsi\phi\KS}$ is defined to be $\pm15 \mev$ around the nominal $\Bd$ mass~\cite{PDG2022}.

\begin{figure}[tb]
\centering
\includegraphics[width=0.47\textwidth]{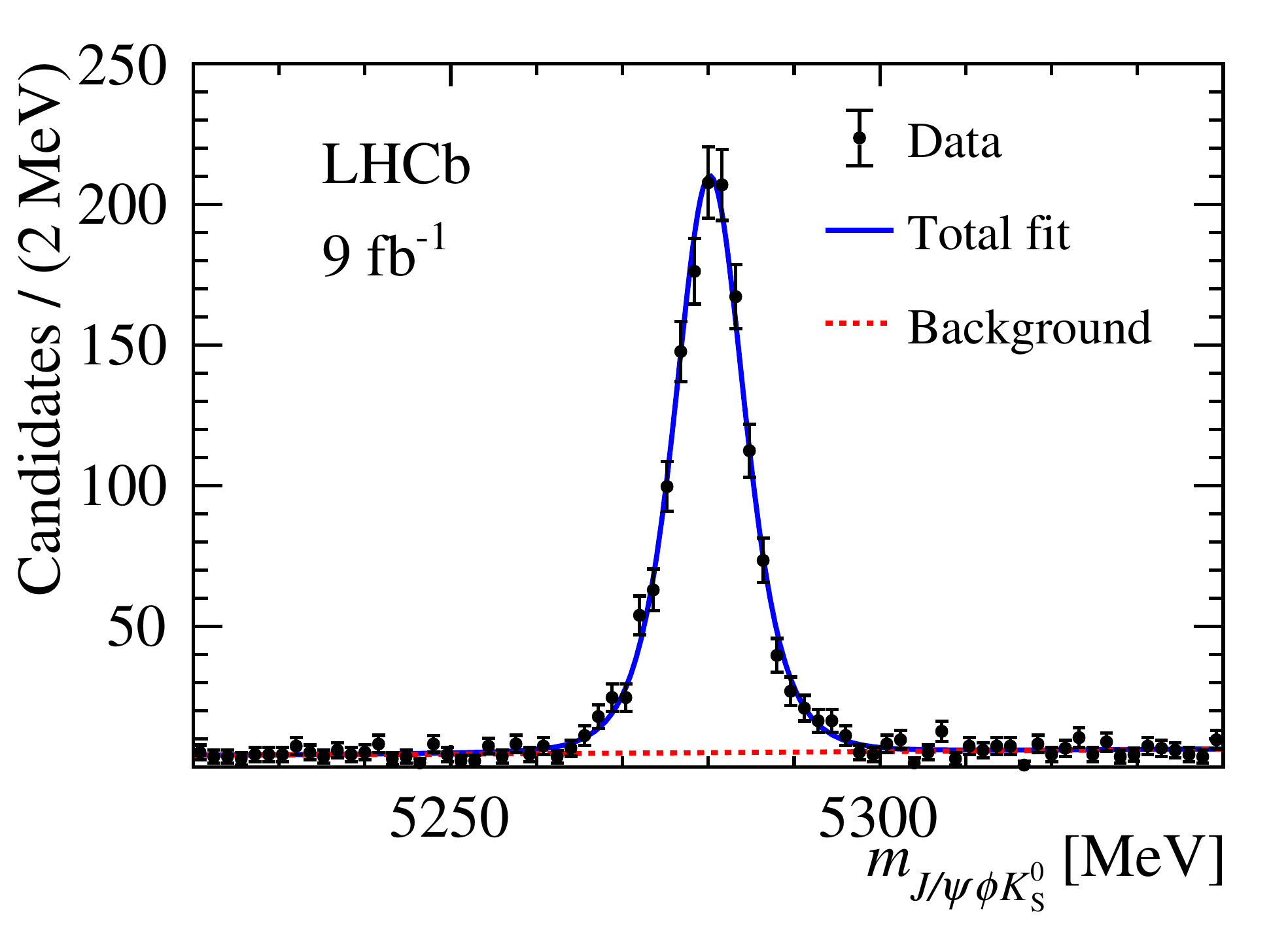}
\put(-32,128){(a)}
\includegraphics[width=0.47\textwidth]{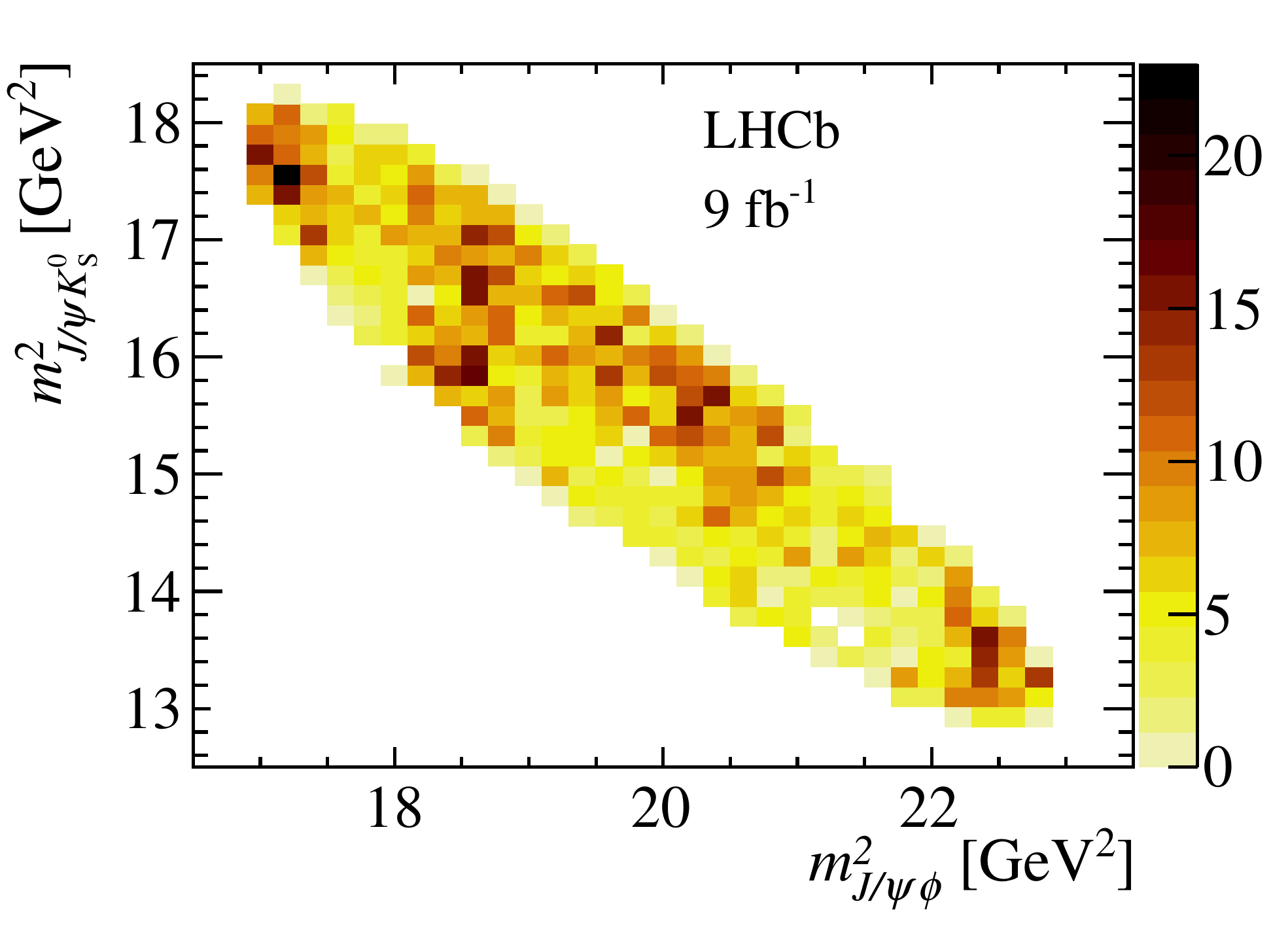}
\put(-50,128){(b)}
\caption{(a) Invariant-mass distribution of selected \Bd candidates and corresponding fit result. (b) Distribution of $m^2_{\jpsi\KS}$ versus $m^2_{\jpsi\phi}$ for candidates in the $\pm$15\mev region around the nominal $\Bz$ mass.}
\label{fig:massfit}
\end{figure}

An extended maximum-likelihood fit is performed to the reconstructed $\Bz$ mass, as is shown in Fig.~\ref{fig:massfit}a. 
The $\Bd$ signal is described by a Hypatia function~\cite{Santos:2013gra} and the background is described by an exponential function.
The $\Bz$ yield is measured to be $1866\pm47$ in the signal region.
The fraction of combinatorial background in the signal region is $6\,\%$.
Roughly $4\,\%$ of the \Bz yield corresponds to peaking background \mbox{$\Bd \to \jpsi \Kp \Km \KS$} decays without an intermediate $\phi$ meson (referred to as non-$\phi$). 
The non-$\phi$ contribution is neglected in the default amplitude model and is considered as a source of systematic uncertainty.
A further kinematic fit is performed to improve the momentum resolution of the final-state particles by constraining the measured $\Bd$ mass to its nominal value.
Figure~\ref{fig:massfit}b shows the distribution of $m^2_{\jpsi\KS}$ versus $m^2_{\jpsi\phi}$ for candidates in the signal region.

An amplitude fit is performed to study the intermediate states. 
The $\Bzdecay$ sample size is small, while the intermediate contributions are complicated, making the fit difficult to converge.
In order to solve this difficulty, a simultaneous fit is performed to the $\Bzdecay$ sample and the $\Bpdecay$ sample in Ref.~\cite{LHCb-PAPER-2020-044}, where the amplitudes in the two decay modes are related through isospin symmetry. 
The likelihood for the $\Bz$ decay is given by
\begin{equation}
\mathcal{L}(\vec{\omega})=
\prod_i\left[ (1-\beta)\, \mathcal{P}_{\rm sig}(m_{\phi \KS}^i,\vec{\Omega}^i|\vec{\omega})
    + \beta\, \mathcal{P}_{\rm bkg}(m_{\phi \KS}^i,\vec{\Omega}^i) \right],\label{eq:likelihood}
\end{equation}  
where $\mathcal{P}_{\rm sig}(m_{\phi \KS}^i,\vec{\Omega}^i|\vec{\omega})$ and $\mathcal{P}_{\rm bkg}(m_{\phi \KS}^i,\vec{\Omega}^i)$ refer to the probability density functions (PDFs) for the signal and background components, respectively.
The superscript $i$ refers to the $i$-th candidate. 
The fraction of combinatorial background, $\beta$, is fixed to $6\,\%$.
The likelihood of the $\Bp$ decay is similar to Eq.~\ref{eq:likelihood}, and is described in detail in Ref.~\cite{LHCb-PAPER-2016-019}.
The decay kinematics are described by a mass, chosen to be $m_{\phi \KS}$, and five angular variables $\vec{\Omega}$ as defined below.
The signal PDF is proportional to the incoherent sum of the squared matrix elements for different final-state muon helicities $(\lambda_\mup,\lambda_\mun)$, and depends on the set of fit parameters, $\vec{\omega}$, which includes masses, widths, and helicity couplings of intermediate states contributing to  $\Bzdecay$ decays,
\begin{equation}
\label{eq:sigpdf}
\mathcal{P}_{\rm sig}(m_{\phi \KS}^{i},\vec{\Omega}^{i}|\vec{\omega})
=\frac{1}{I(\vec{\omega})}\sum_{\lambda_\mup\lambda_\mun}\sum_{f}{\big|\mathcal{M}_{f}(m_{\phi \KS}^i,\vec{\Omega}^i| \vec{\omega})\big|^2}
\Phi(m_{\phi \KS}^i, \vec{\Omega}^i)
\epsilon(m_{\phi \KS}^i,\vec{\Omega}^i). 
\end{equation}
The term $\Phi(m_{\phi \KS}^i, \vec{\Omega}^i)$ represents the phase-space density, the term $\epsilon(m_{\phi \KS}^i,\vec{\Omega}^i)$  represents the efficiency, and the term $I(\vec{\omega})$ is a normalisation factor.
The efficiency is obtained from simulated samples.
Since the $\Bzdecay$ decay is a self-charge-conjugated mode, the data sample contains both $\Bz$ and $\Bzb$ decays with unknown $\bquark$ flavour. Therefore, the signal PDF for the neutral $B$ decay averages $\Bz$ and $\Bzb$ contributions, where the flavour of $B$ candidates is denoted as $f$ in Eq.~\ref{eq:sigpdf}.

The matrix element $\mathcal{M}_f(m_{\phi \KS}^i,\vec{\Omega}^i| \vec{\omega})$ is constructed based on the helicity formalism~\cite{Chung:186421,*Richman:1984gh,*Jacob:1959at}. 
Three interfering decay sequences are considered: \mbox{$\decay{\Bz}{(K^{*}\to\phi \KS)\jpsi}$}, \mbox{$\decay{\Bz}{(X\to\jpsi\phi) \KS}$}, and \mbox{$\decay{\Bz}{(\zcsAll \to\jpsi \KS)\phi}$}.\footnote{The $\Kstar$ denotes any excited $K$ state. The $X$ denotes $\chi_{c0,1}$, $\eta_{c2}$, and $T_{\psi\phi1}^{\eta}$ states.} 
The decay sequences can be described by $m_{\phi\KS}$ and $\vec{\Omega}\equiv( \theta_{K^{*}}, \theta_{\jpsi}, \theta_{\phi}, \Delta\varphi_{K^{*}\jpsi}, \Delta\varphi_{K^{*}\phi})$, where the $\theta_{\Kstar}$, $\theta_{\jpsi}$, and $\theta_{\phi}$ are the helicity angles of the $\phi$, $\mu$, and $\Kp$ particles in the $\Kstar$, $\jpsi$, and $\phi$ rest frames, respectively.
The angles between the $K^{*}$ and $\jpsi$ decay planes, and between the $\Kstar$ and $\phi$ decay planes in the $B$ rest frame, are denoted as $\Delta \varphi_{K^{*}\jpsi}$, and $\Delta \varphi_{K^{*}\phi}$, respectively. 
The background PDF $\mathcal{P}_{\rm bkg}(m_{\phi \KS}^i,\vec{\Omega}^i)$ is determined 
from studying the candidates in the sideband of the $m_{\jpsi\phi\KS}$ distribution with the methods outlined in Ref.~\cite{LHCb-PAPER-2016-019}.

The default model is taken from Ref.~\cite{LHCb-PAPER-2020-044} and is composed of 
nine $\Kstar$ components, seven $X$ components, and two \zcsAll states ($\zcsOne$ and $\zcsTwo$), and one $\jpsi\phi$ nonresonant component (NR).
Relativistic Breit--Wigner (RBW) functions are used to model the lineshapes of these resonances. 
The amplitudes of the $\Bz$ and the $\Bp$ decays are formed by intermediate states that are either identical (namely $X$) or isospin partners ($\Kstar$ or \zcsAll). 
Under the assumption of isospin symmetry, the mass, width, and helicity couplings for all the components except for the $\zcsOne$ states are constrained to be identical. 
In order to test the significance of the \zcsz signal without prior knowledge of its properties, a fit is performed with the masses, widths, and helicity couplings of \zcsz and \zcsp states allowed to vary independently. 
In an alternative model, isospin symmetry is imposed for the \zcsz and \zcsp  components.
The parameters for the $\zcsTwo^{0}$ state are always constrained to be identical to the $\zcsTwo^{+}$ state due to the limited size of the $\Bz$ sample.

\begin{figure}[tb]
\centering
\includegraphics[width=1.0\textwidth]{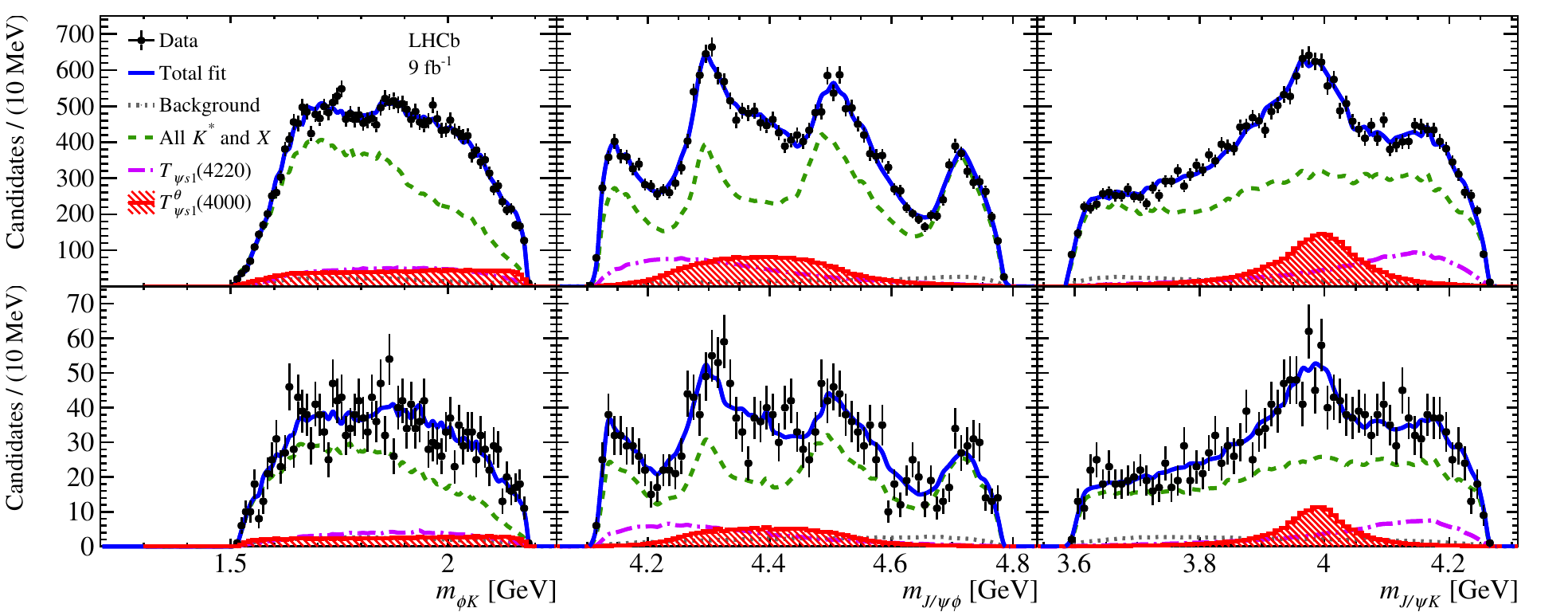}%
\caption{
Distributions of (left) $m_{\phi K}$, (middle) $m_{\jpsi\phi}$, and (right) $m_{\jpsi K}$, overlaid with the corresponding projections of the default fit model. 
The upper and lower rows correspond to the $\Bpdecay$ and $\Bzdecay$ decays, respectively.
}
\label{fig:fitmjpsik}
\end{figure}

Figure \ref{fig:fitmjpsik} shows the $\phi K$, $\jpsi\phi$, and $\jpsi K$ mass distributions and the corresponding fit projections of the default model for the two $B$ decay modes.
Table~\ref{tab:fit_results_of_zcs_4000_conclusion} summarises measurements of the  mass, width, fit fraction of the \zcsz state, and the mass difference between the \zcsz and \zcsp states, defined as \mbox{$\Delta M \equiv M_{\zcsz}-M_{\zcsp}$}. 
The fit value of $\Delta M$ is zero within uncertainties,  consistent with the two states being isospin partners. 
The fit fraction of each component is defined as the integral of the signal PDF divided by the $I(\vec{\omega})$ term.
All the fit parameters, including mass, width, and helicity couplings of the intermediate states, of the default model in this analysis are consistent with the corresponding parameters of the default model for the $\Bpdecay$ decay in Ref.~\cite{LHCb-PAPER-2020-044}.

\begin{table}[bt]
\small
\centering
\caption{Results for the \zcsz state from the default model. The first uncertainty is statistical and the second systematic.}
\label{tab:fit_results_of_zcs_4000_conclusion}
\renewcommand\arraystretch{1.4}
\begin{tabular}{ccccc}
\hline
State  & Mass (MeV)     &  Width (MeV)    &  Fit fraction (\%) & $\Delta M$ (MeV)\\ \hline
\zcsz & $3991  \,_{-10}^{+12} \,^{\phz+9}_{-17}$  &   $105 \,_{-25}^{+29}\,^{+17}_{-23}$  & $7.9\pm{2.5}\,^{+3.0}_{-2.8}$ & $-12 \,_{-10}^{+11} \,^{+6}_{-4}$\\
\hline
\end{tabular}
\end{table}

\begin{table}[tb]
\footnotesize
\renewcommand\arraystretch{1.2}
\begin{center}
\caption{Systematic uncertainties associated to the mass, width, fit fraction  of the \zcsz state, and the mass difference between the \zcsz and \zcsp states.}
\label{tab:sys_all}
\begin{tabular}{ccccc}
\hline

Source              &Mass (\mev)               &Width (\mev)         & Fit fraction (\%)          &  $\Delta M$ (\mev)  \\  
\hline
Efficiency and background models & \fontsize{7.01236}{0}{$+10.46$} & \fontsize{7.01236}{0}{$+\phz1.91$} & \fontsize{7.01236}{0}{$-0.06$} & \fontsize{7.01236}{0}{$+1.02$} \\
Hadron radius           & $_{-\phz1.92}^{+\phz0.29}$ & $_{-\phz5.85}^{+\phz2.33}$ &$_{-0.39}^{+0.05}$ & $_{-1.68}^{+0.92}$ \\
$X(4140)$ Flatt\'e model                           & \fontsize{7.01236}{0}{$+\phz0.61$} & \fontsize{7.01236}{0}{$-\phz3.47$} & \fontsize{7.01236}{0}{$-0.31$} & \fontsize{7.01236}{0}{$-0.18$} \\
$\jpsi \phi$ NR representation                 & \fontsize{7.01236}{0}{$+\phz1.72$} & \fontsize{7.01236}{0}{$+\phz1.46$} & \fontsize{7.01236}{0}{$+0.38$}  & \fontsize{7.01236}{0}{$+0.12$} \\
Simplified K-Matrix model                        & \fontsize{7.01236}{0}{$-\phz7.30$} & \fontsize{7.01236}{0}{$-17.54$} & \fontsize{7.01236}{0}{$-1.73$} & \fontsize{7.01236}{0}{$-3.38$} \\
RBW widths                             & \fontsize{7.01236}{0}{$+\phz1.90$} & \fontsize{7.01236}{0}{$-\phz3.41$} & \fontsize{7.01236}{0}{$-0.56$} & \fontsize{7.01236}{0}{$+1.70$} \\
Spin-parity of the \zcsTwo state       & \fontsize{7.01236}{0}{$-16.32$} & \fontsize{7.01236}{0}{$-17.76$} & \fontsize{7.01236}{0}{$-2.33$} & \fontsize{7.01236}{0}{$-1.48$} \\
Non-$\phi$ contribution                          & \fontsize{7.01236}{0}{$+\phz5.52$} & \fontsize{7.01236}{0}{$+\phz7.98$} & \fontsize{7.01236}{0}{$+0.92$}  & \fontsize{7.01236}{0}{$+5.49$} \\
Additional $1^+$ $\jpsi \phi$ NR                 & \fontsize{7.01236}{0}{$+\phz0.80$}  & \fontsize{7.01236}{0}{$-\phz3.66$} &\fontsize{7.01236}{0}{$+0.18$} & \fontsize{7.01236}{0}{$-1.28$} \\
Additional $2^+$ $\jpsi \phi$ NR                 & \fontsize{7.01236}{0}{$+\phz7.66$}  & \fontsize{7.01236}{0}{$+\phz2.57$} &\fontsize{7.01236}{0}{$+2.52$} & \fontsize{7.01236}{0}{$+0.92$} \\
Extended model                                   & \fontsize{7.01236}{0}{$-\phz6.36$}  & \fontsize{7.01236}{0}{$-\phz2.54$} &\fontsize{7.01236}{0}{$-0.20$} & \fontsize{7.01236}{0}{$-2.98$} \\
Additional $X$ state                             & \fontsize{7.01236}{0}{$-\phz5.58$}  & \fontsize{7.01236}{0}{$+\phz0.27$} &\fontsize{7.01236}{0}{$+0.34$} & \fontsize{7.01236}{0}{$-1.48$} \\ 
Fit fraction constraint                          & \fontsize{7.01236}{0}{$+\phz0.07$} &  \fontsize{7.01236}{0}{$+\phz1.94$} & \fontsize{7.01236}{0}{$-0.48$} & \fontsize{7.01236}{0}{$-0.45$}\\
\hline
Impact parameter $\chi^{2}$ modelling         & \fontsize{7.01236}{0}{$+\phz1.00$}  & \fontsize{7.01236}{0}{$-\phz2.49$} & \fontsize{7.01236}{0}{$-0.22$} & \fontsize{7.01236}{0}{$+0.72$}  \\   
\hline
Finite simulation sample size                     & \fontsize{7.01236}{0}{$+\phz2.90$} & \fontsize{7.01236}{0}{$+11.20$}  &  \fontsize{7.01236}{0}{$+1.33$} & \fontsize{7.01236}{0}{$+2.17$} \\
\hline
Fixed mass and width of $\Kstar$ states     & \fontsize{7.01236}{0}{$-\phz0.58$} & \fontsize{7.01236}{0}{$-\phz1.30$} & \fontsize{7.01236}{0}{$-0.43$} & \fontsize{7.01236}{0}{$-0.58$}  \\
\hline
Samples to construct the $\mathcal{P}_{\rm bkg}$ & $_{-\phz1.38}^{+\phz1.06}$ & $_{-\phz7.25}^{+\phz6.38}$ & $_{-0.59}^{+0.54}$ & $_{-0.68}^{+0.52}$\\
\hline
Background fraction uncertainty  & $_{-\phz0.45}^{+\phz0.37}$ & $_{-\phz1.55}^{+\phz1.38}$ &$_{-0.07}^{+0.06}$ & $_{-0.38}^{+0.22}$ \\
\hline
\textbf{Final}                          & $_{-16.71}^{+\phz8.46}$   & $_{-23.33}^{+17.09}$      & $_{-2.84}^{+3.00}$   &$_{-4.23}^{+6.04}$ \\
\hline
\end{tabular}
\end{center}
\end{table}

The estimated systematic uncertainties on the mass, width, fit fraction of the \zcsz state, and on $\Delta M$ are summarised in Table~\ref{tab:sys_all}.
Both the background PDF and efficiency function are described by an expansion with Legendre polynomials and a spherical harmonic function instead of interpolation. 
The Blatt–Weisskopf hadron size is varied between $1.5$ and $4.5 \gev^{-1}$ from the default value $3\gev^{-1}$. 
The Flatt\'e model~\cite{Flatte:1976xu} is used for the lineshape of the $X(4140)$ state instead of the RBW function to account for the possible additional decay channel $\Dssp\Dsm$.
The $\jpsi\phi$ NR shape is changed from constant to be a linear function of $m_{\jpsi\phi}$.  
The $\Kstar$ states are described by a simplified one-channel K-matrix model~\cite{PDG2022} instead of a sum of RBW functions.
The mass-dependent widths in the RBW function for the $\Kstar$ resonances are calculated using the decay with the largest branching fraction, $\Kstar\to K \pi$ or $\Kstar\to\Kstar(892)\pi$, instead of using the $\Kstar\to\phi K$ decay. 
The spin-parity of the \zcsTwo state is changed from the default $1^+$ to $1^-$ to account for its ambiguity between $1^+$ and $1^-$~\cite{LHCb-PAPER-2020-044}.  
The uncertainty originating from the neglected non-$\phi$ contribution is evaluated by narrowing the $\phi$ mass window from $\pm15\mev$ to $\pm8\mev$. 
An additional $\jpsi\phi$ NR contribution with the spin-parity equal to $1^+$ or $2^+$ is included.
An extended model with more $\Kstar$ states is also studied, which includes all the $\Kstar$ resonances that are within the allowed phase space, as predicted in Ref.~\cite{Godfrey:1985xj}.
Possible additional $X$ states, with spins ranging from 0 to 2, are checked based on the extended model.
The total fit fraction of the default model is 165.2\,\%, which indicates that the interference between the decay amplitudes is large. An alternative fit constraining the total fit fraction to be smaller than 140\,\%, corresponding to a reduction by three times its uncertainty, is performed. 
The final positive (negative) systematic uncertainty is taken as the maximal positive (negative) deviation from the model uncertainties above summed in quadrature with the uncertainty from other sources.
These other sources of systematic uncertainty include 
mis-modelling of the impact parameter $\chi^{2}$ of $B$ candidates, 
the finite size of the simulated samples, 
the fixed masses and widths of the known $\Kstar$ states, 
the choice of the data sample used to construct the background PDF model, 
and the uncertainty on the background fraction. 

The significance of the \zcsz state is evaluated with a likelihood ratio test.
The test statistic is defined as  $t \equiv -2\ln[\mathcal{L}(H_0)/\mathcal{L}(H_1)]$, 
with $H_1$ and $H_0$ denoting the default model with and without the \zcsz state.
Here, the term $\mathcal{L}$ represents the likelihood of the $\Bz$ decay.
Five thousand pseudoexperiments are generated with the $H_0$ model, where the parameters of the model are fixed to the values determined from data.
The sample size of each pseudoexperiment follows a Poisson distribution with its mean being equal to the number of candidates in the data sample.
A fit is performed to each pseudoexperiment with the $H_0$ and $H_1$ models to determine the test statistic $t$.
The $t$ value from these pseudoexperiments should follows a $\chi^2$ distribution, where the number of degrees-of-freedom is a freely varying parameter. 
Using the resulting $\chisq$ distribution, the probability of $t > t_{\rm data}$ is taken as the $p$-value.
The corresponding  significance is estimated to be $4.9\sigma$, decreasing to $4.0\sigma$ after including systematic uncertainties, which provides evidence for the \zcsz state.
The significance for the $\zcsz$ state is taken as the smallest significance found when varying the sources of systematic uncertainty.

The significance is also evaluated assuming isospin symmetry for the $\zcsOne$ states, 
 where the test statistic is defined as $t^{\prime} \equiv -2\ln[\mathcal{L}(H_0)/\mathcal{L}(H_1^{\prime})]$, with $\mathcal{L}$ denoting the total likelihood of the $\Bp$ and $\Bz$ decays.
The corresponding fit parameters for the \zcsp and \zcsz states, including mass, width, and helicity couplings, are constrained to be equal between the two states in the $H_1^{\prime}$ model. 
A fit is performed to the $t^{\prime}$ distribution from the pseudoexperiments with a Gaussian function. 
The Gaussian function rather than a $\chisq$ distribution is used here because the number of degree-of-freedoms of the $H_0$ and $H_1^{\prime}$ models are equal.
The significance is estimated to be $7.2\sigma$ and decreases to $5.4\sigma$ after accounting for systematic uncertainties.

In conclusion, an amplitude analysis of the $\Bd \to \jpsi \phi \KS$ decay is performed.
Evidence of a $\jpsi \KS$ structure, denoted the \zcsz state, is obtained with a significance of $4.0\sigma$.
The mass and width of this state are measured to be 
\begin{equation}
\begin{aligned}
M(\zcsz)      &= 3991 \,_{-10}^{+12} \,^{+\phz9}_{-17} \mev, \\
\Gamma(\zcsz) &= \phz105  \,_{-25}^{+29}\,^{+17}_{-23} \mev,
\nonumber
\end{aligned}
\end{equation}
where the first uncertainty is statistical and the second systematic.
The mass difference between the \zcsz and \zcsp states is measured to be
\begin{equation}
\begin{aligned}
\Delta M = -12 \,_{-10}^{+11}\,^{+6}_{-4} \mev,
\nonumber
\end{aligned}
\end{equation}
which is consistent with the two states being isospin partners.
With isospin symmetry imposed for the $\zcsOne$ states, 
the significance of the \zcsz structure is measured to be $5.4\sigma$.

\section*{Acknowledgements}
%
%
\noindent We express our gratitude to our colleagues in the CERN
accelerator departments for the excellent performance of the LHC. We
thank the technical and administrative staff at the LHCb
institutes.
We acknowledge support from CERN and from the national agencies:
CAPES, CNPq, FAPERJ and FINEP (Brazil); 
MOST and NSFC (China); 
CNRS/IN2P3 (France); 
BMBF, DFG and MPG (Germany); 
INFN (Italy); 
NWO (Netherlands); 
MNiSW and NCN (Poland); 
MEN/IFA (Romania); 
MICINN (Spain); 
SNSF and SER (Switzerland); 
NASU (Ukraine); 
STFC (United Kingdom); 
DOE NP and NSF (USA).
We acknowledge the computing resources that are provided by CERN, IN2P3
(France), KIT and DESY (Germany), INFN (Italy), SURF (Netherlands),
PIC (Spain), GridPP (United Kingdom), 
CSCS (Switzerland), IFIN-HH (Romania), CBPF (Brazil),
Polish WLCG  (Poland) and NERSC (USA).
We are indebted to the communities behind the multiple open-source
software packages on which we depend.
Individual groups or members have received support from
ARC and ARDC (Australia);
Minciencias (Colombia);
AvH Foundation (Germany);
EPLANET, Marie Sk\l{}odowska-Curie Actions and ERC (European Union);
A*MIDEX, ANR, IPhU and Labex P2IO, and R\'{e}gion Auvergne-Rh\^{o}ne-Alpes (France);
Key Research Program of Frontier Sciences of CAS, CAS PIFI, CAS CCEPP, 
Fundamental Research Funds for the Central Universities, 
and Sci. \& Tech. Program of Guangzhou (China);
GVA, XuntaGal, GENCAT and Prog.~Atracci\'on Talento, CM (Spain);
SRC (Sweden);
the Leverhulme Trust, the Royal Society
 and UKRI (United Kingdom).


\ifx\mcitethebibliography\mciteundefinedmacro
\PackageError{LHCb.bst}{mciteplus.sty has not been loaded}
{This bibstyle requires the use of the mciteplus package.}\fi
\providecommand{\href}[2]{#2}

\clearpage
\centerline
{\large\bf LHCb collaboration}
\begin
{flushleft}
\small
R.~Aaij$^{32}$\lhcborcid{0000-0003-0533-1952},
A.S.W.~Abdelmotteleb$^{50}$\lhcborcid{0000-0001-7905-0542},
C.~Abellan~Beteta$^{44}$,
F.~Abudin{\'e}n$^{50}$\lhcborcid{0000-0002-6737-3528},
T.~Ackernley$^{54}$\lhcborcid{0000-0002-5951-3498},
B.~Adeva$^{40}$\lhcborcid{0000-0001-9756-3712},
M.~Adinolfi$^{48}$\lhcborcid{0000-0002-1326-1264},
P.~Adlarson$^{77}$\lhcborcid{0000-0001-6280-3851},
H.~Afsharnia$^{9}$,
C.~Agapopoulou$^{13}$\lhcborcid{0000-0002-2368-0147},
C.A.~Aidala$^{78}$\lhcborcid{0000-0001-9540-4988}
% ...

{\footnotesize \it

% ...
$^{9}$Universit{\'e} Clermont Auvergne, CNRS/IN2P3, LPC, Clermont-Ferrand, France\\
% ...
$^{13}$LPNHE, Sorbonne Universit{\'e}, Paris Diderot Sorbonne Paris Cit{\'e}, CNRS/IN2P3, Paris, France\\
% ...
$^{32}$Nikhef National Institute for Subatomic Physics, Amsterdam, Netherlands\\
% ...
$^{40}$Instituto Galego de F{\'\i}sica de Altas Enerx{\'\i}as (IGFAE), Universidade de Santiago de Compostela, Santiago de Compostela, Spain\\
% ...
$^{44}$Physik-Institut, Universit{\"a}t Z{\"u}rich, Z{\"u}rich, Switzerland\\
% ...
$^{48}$H.H. Wills Physics Laboratory, University of Bristol, Bristol, United Kingdom\\
% ...
$^{50}$Department of Physics, University of Warwick, Coventry, United Kingdom\\
% ...
$^{53}$School of Physics and Astronomy, University of Glasgow, Glasgow, United Kingdom\\
$^{54}$Oliver Lodge Laboratory, University of Liverpool, Liverpool, United Kingdom\\
% ...
$^{62}$Syracuse University, Syracuse, NY, United States\\
% ...
$^{77}$Department of Physics and Astronomy, Uppsala University, Uppsala, Sweden, associated to $^{53}$\\
$^{78}$University of Michigan, Ann Arbor, MI, United States, associated to $^{62}$\\
\bigskip
% ...

}
\end{flushleft}

\end{document}